\documentclass[3p]{elsarticle}
\usepackage{amsmath,amssymb,amsfonts}
\usepackage{algorithmic}
\usepackage{graphicx}
\usepackage{textcomp}
\usepackage{xcolor}
\usepackage{wrapfig}
\usepackage{balance}

\usepackage{booktabs}
\usepackage{multirow}
\usepackage[T1]{fontenc} 
\usepackage[utf8]{inputenc}
\usepackage[english]{babel} 
\PassOptionsToPackage{hyphens}{url}\usepackage{hyperref}
\usepackage{lipsum}
\usepackage{tabularx}

\newcommand{\fullstack}{\textsc{FullStackDeveloper}}
\newcommand{\backend}{\textsc{BackendDeveloper}}
\newcommand{\sysadmin}{\textsc{SystemAdministrator}}
\newcommand{\frontend}{\textsc{FrontendDeveloper}}
\newcommand{\mobile}{\textsc{MobileDeveloper}}
\newcommand{\qatest}{\textsc{QATestDeveloper}}
\newcommand{\devops}{\textsc{DevOpsDeveloper}}
\newcommand{\databaseadmin}{\textsc{DatabaseAdministrator}}
\newcommand{\desktop}{\textsc{DesktopDeveloper}}
\newcommand{\datasicentist}{\textsc{DataScientist}}
\newcommand{\embedded}{\textsc{EmbeddedDeveloper}}
\newcommand{\productmanager}{\textsc{ProductManager}}
\newcommand{\gamedeveloper}{\textsc{GameDeveloper}}
\newcommand{\designer}{\textsc{Designer}}

\newcommand{\datasystems}{\textit{Data Systems}} 
\newcommand{\devtools}{\textit{Development Tools}} 
\newcommand{\languages}{\textit{Languages}} 
\newcommand{\libsframeworks}{\textit{Libs \& Frameworks}} 
\newcommand{\osinfrastructure}{\textit{OS \& Infrastructure}} 
\newcommand{\processmethods}{\textit{Process \& Methods}}

\let\ElseVierBibliography\bibliography%
\renewcommand{\bibliography}[1]{%
\section*{References}%
\ElseVierBibliography{#1}%
}%

\graphicspath{{figs/}}

\def\BibTeX{{\rm B\kern-.05em{\sc i\kern-.025em b}\kern-.08em
    T\kern-.1667em\lower.7ex\hbox{E}\kern-.125emX}}

\begin{document}

\title{
    What Skills do IT Companies look for in New Developers? \\
    {A Study with Stack Overflow Jobs}
}

\author[ufmg]{João Eduardo Montandon\corref{correspondingauthor}}
\cortext[correspondingauthor]{Corresponding author}
\ead[url]{joao.montandon@dcc.ufmg.br}

\author[concordia]{Cristiano Politowski}
\author[ifmg]{Luciana Lourdes Silva}
\author[ufmg]{Marco Tulio Valente}
\author[chicoutimi]{Fabio Petrillo}
\author[concordia]{Yann-Ga\"el Gu\'{e}h\'{e}neuc}

\address[ufmg]{Federal University of Minas Gerais}
\address[concordia]{Concordia University}
\address[ifmg]{Federal Institute of Minas Gerais}
\address[chicoutimi]{Université du Québec à Chicoutimi}


\begin{abstract}
    \textbf{Context:} There is a growing demand for information on how IT companies look for candidates to their open positions.
    \textbf{Objective:} This paper investigates which hard and soft skills are more required in IT companies by analyzing the description of 20,000 job opportunities.
    \textbf{Method:} We applied open card sorting to perform a high-level analysis on which types of hard skills are more requested.
    Further, we manually analyzed the most mentioned soft skills.
    \textbf{Results:} Programming languages are the most demanded hard skills. 
    Communication, collaboration, and problem-solving are the most demanded soft skills.
    \textbf{Conclusion:} We recommend developers to organize their resumé according to the positions they are applying. 
    We also highlight the importance of soft skills, as they appear in many job opportunities.
\end{abstract}


\maketitle


Software is ``eating the world'' and we everyday observe the rise of companies centered on software.
At the same time, software is becoming more complex, with its development requiring highly qualified developers~\cite{Montandon2019}.
As a result, modern IT companies allocate developers to work in specific areas, such as databases, security, user interface (front-end design), core features (back-end design), mobile development, etc.

Therefore, there is a growing demand for information on how modern companies deal with this context when looking for developers to their open positions.
To find out what are the skills required by companies when selecting new employees, we report in this article the first results of an analysis of more than 20,000 job opportunities available in Stack Overflow Jobs portal, a platform that allows companies to publish new opportunities for IT professionals.
In particular, we leveraged the main characteristics of 14 IT roles, such as Backend, Frontend, and Mobile developers.
We also analyze which soft skills are mostly requested by IT companies when selecting new candidates. 
Finally, we discuss the implications of this analysis for both recruiters and developers.

\section{The Anatomy of a Job Opportunity}

\begin{figure*}[t!]
    \centering
    \includegraphics[width=.85\textwidth]{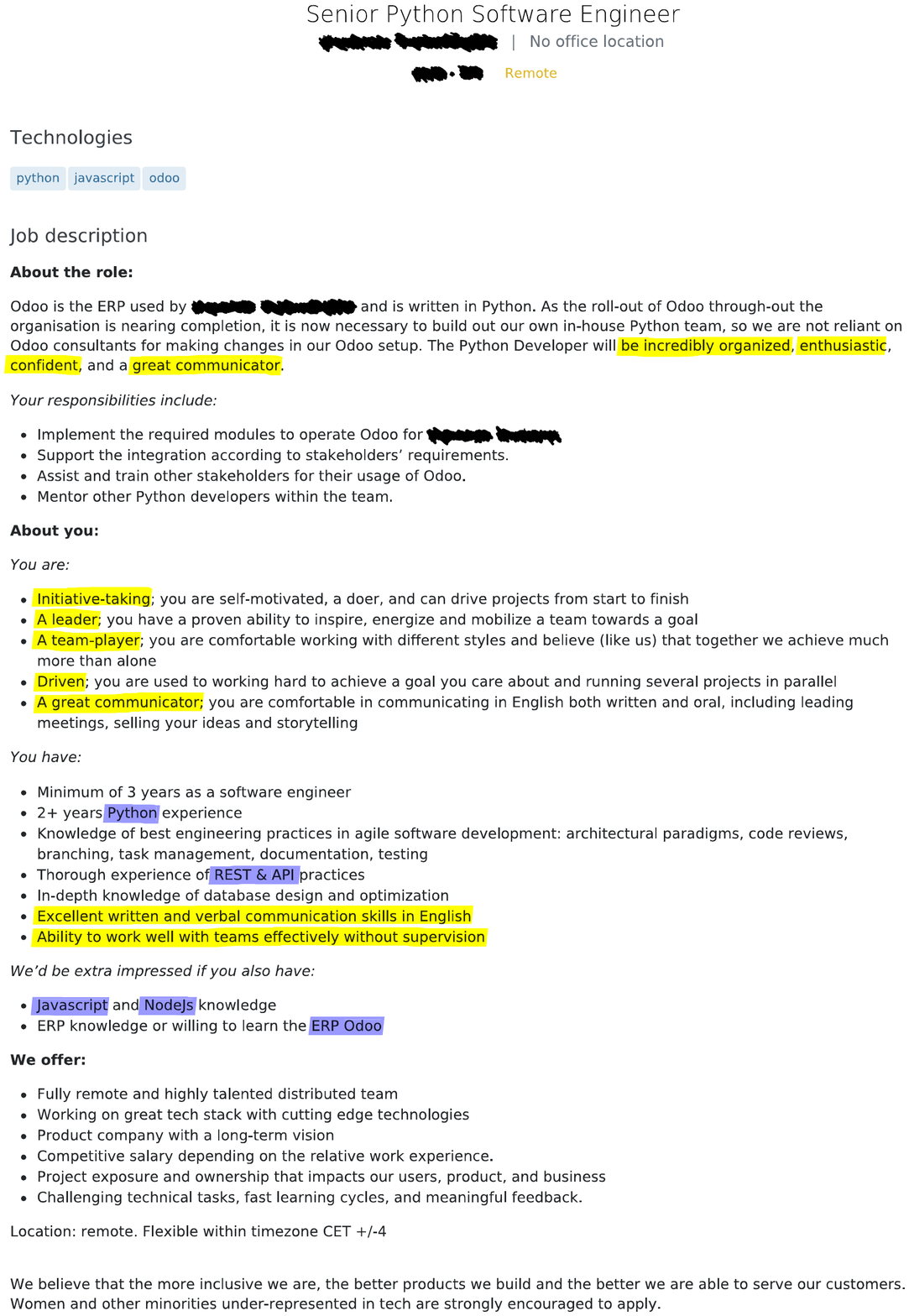}
    \caption{An example of a post in Stack Overflow Jobs portal. Hard skills are annotated in blue, while soft skills are in yellow.}
    \label{fig:job-post}
\end{figure*}

A job opportunity is a declaration of expectations.
It describes what a company expects from its candidates, as well as what the candidates should expect from the company.
For this, a job opportunity includes important company details, such as mission, culture, benefits, etc.
At the same time, the company should provide enough details so candidates can decide whether they are qualified or not for the position.
Lastly, an opportunity should list the job's responsibilities, so that candidates can be aware of their duties.

\autoref{fig:job-post} shows an example of a job opportunity posted on Stack Overflow Jobs portal.
As we can observe, some skills are technically-driven, e.g., \textit{Python}, \textit{REST \& API}, \textit{JavaScript}, etc.
Skills in this group are known as hard skills~\citep{Xia2019}.
By contrast, other skills denote behavioral characteristics of the candidates, such as \textit{verbal communication}, \textit{team player}, \textit{leadership}, etc.
They are known as soft skills~\citep{Sayfullina2018}.
In this article, we analyze both hard and soft skills required by IT job opportunities.

\section{Data Collection}
\label{sec:data-collection}

We investigated the jobs posts available on Stack Overflow Jobs portal.\footnote{\url{https://stackoverflow.com/jobs}} 
We collected the posts visible in the platform for three months, from March 25\textsuperscript{th} to June 28\textsuperscript{th}, 2019 by downloading the posts available in every weekday during this period.
As a result, we retrieved a total of 20,968 job posts, including their titles, description, and tags.
Furthermore, each post in Stack Overflow is automatically associated with at least one out of 14 predefined roles provided by the platform, such as \backend, \frontend, \mobile, etc.
We also collected this information to perform a fine-grained analysis of each post.

\begin{figure}[t!]
    \centering
    \includegraphics[width=0.8\linewidth]{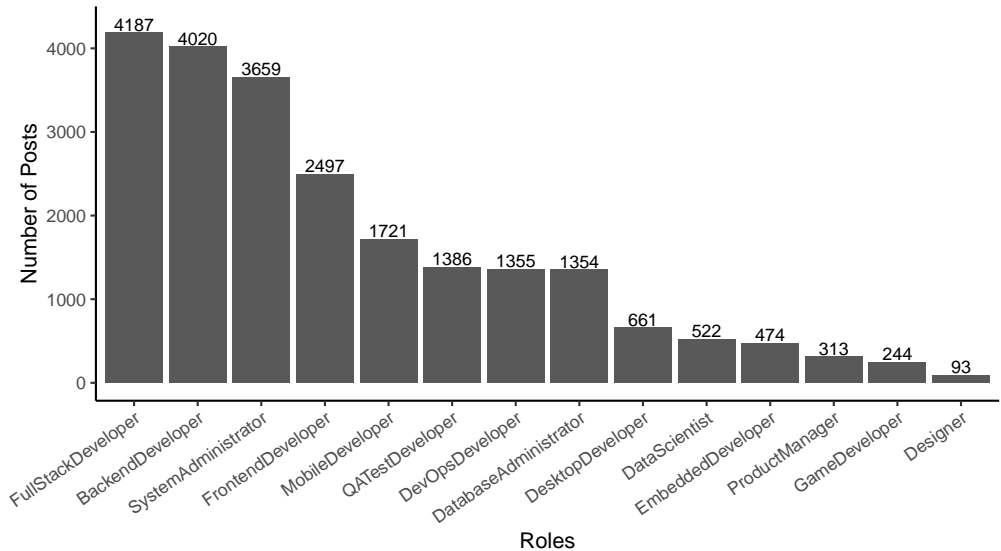}
    \caption{Distribution of job posts collected in this study.}
    \label{fig:roles-barchart}
\end{figure}

\autoref{fig:roles-barchart} shows the distribution of opportunities for each role.
\fullstack, \backend, and \sysadmin\ are the most demanded positions, with at least 3,600 posts each, representing 52.8\% of all job posts.
By contrast, six roles have less than 1,000 posts: \desktop, \datasicentist, \embedded, \productmanager, \gamedeveloper, and \designer.
Together they represent 10.2\% of the analyzed posts.

\section{Analyzing Hard Skills}
\label{sec:hard-skills}

On Stack Overflow Jobs, companies describe the hard skills required for a given position using the same tagging mechanism provided by Stack Overflow's main platform.
Tags play a central role in this ecosystem, as they are used to identify Q\&A topics~\citep{Saha2013}.
Therefore, we use the tags added to each post as an indication of the hard skills demanded by each one.
For instance, the tags associated to the job post in \autoref{fig:job-post} (top) indicate that candidates should master \textit{python}, \textit{javascript}, \textit{NodeJS}, and \textit{odoo}. 

In total, we identified 1,916 hard skills (i.e.,~tags) mentioned 70,680 times in the collected posts.
The frequency of these skills follows a long tail shape, where few elements are generally responsible for almost all occurrences.
For this reason, we removed the hard skills with less than 10 occurrences.
We ended up with 282 hard skills, representing 67,062 occurrences (95\%).
Next, two authors used open card sorting~\citep{Spencer2009} to group tags into abstract categories.
After analyzing together a group of 50 tags, they leveraged six abstract hard skills: \languages, \libsframeworks, \osinfrastructure, \processmethods, \datasystems, and \devtools.
Then, they independently annotated the remaining 232 tags into these categories  (i.e., each tag was evaluated by two authors).
The inter-rater agreement, calculated using Cohen's Kappa~\citep{Kappa}, was 0.82.
Lastly, both authors discussed the conflicting cases to reach a common ground.

\begin{figure}[t!]
    \centering
    \includegraphics[width=0.9\columnwidth]{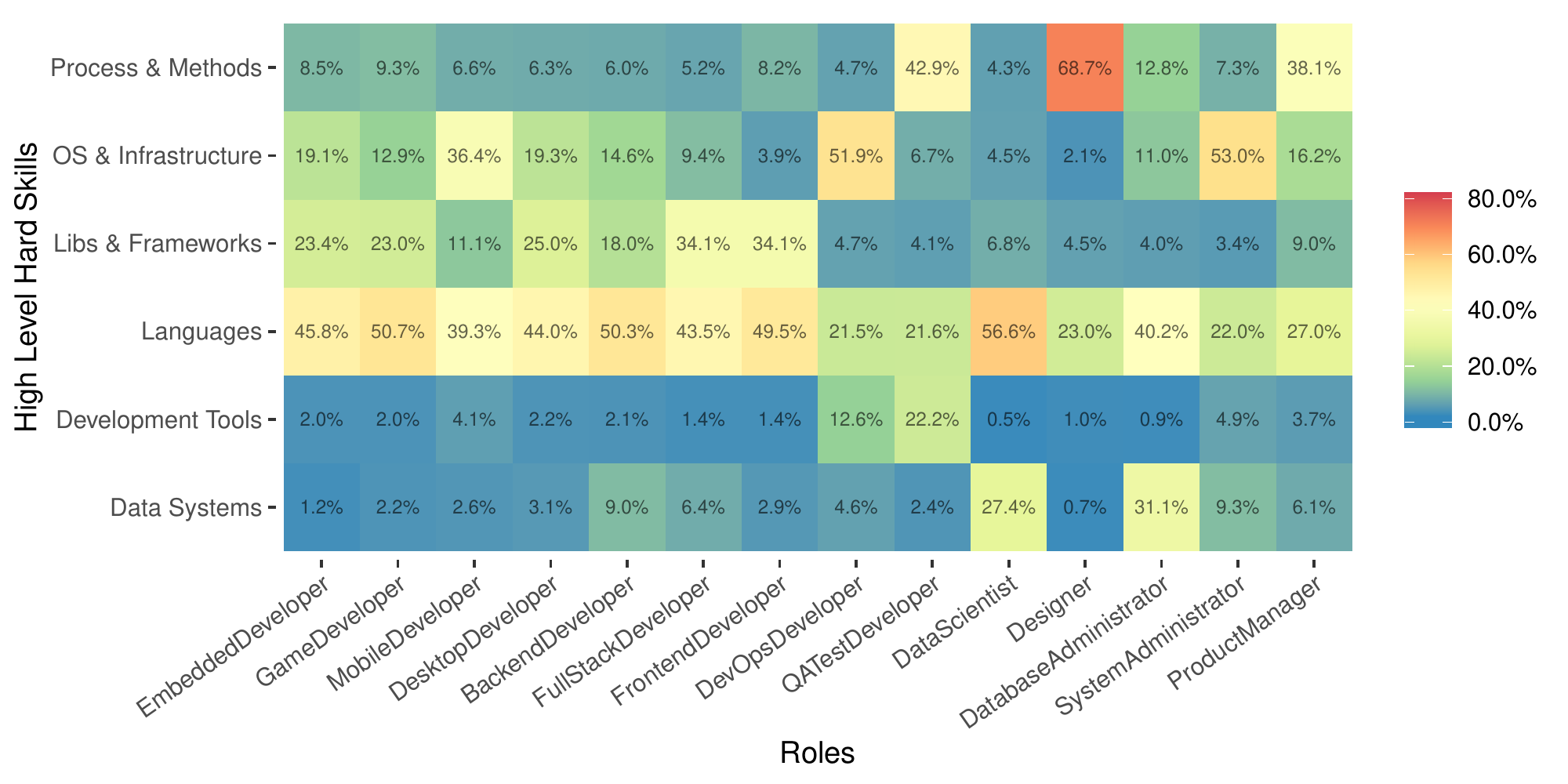}
    \caption{Heatmap indicating how hard skills are distributed among developers roles.}
    \label{fig:heatmap-roles-hs-categories}
\end{figure}

\autoref{fig:heatmap-roles-hs-categories} shows a heatmap with the distribution of high-level hard skills (rows) per developer roles (columns).
With this heatmap, we intend to provide a technology-agnostic analysis, i.e., one that focuses on high-level hard skills (e.g., \languages) instead of current technologies (e.g., \textit{Java}).
In this way, we also intend to increase the validity of our results against technological changes.

First, \languages-based skills are relevant for all roles, ranging from 21.5\% (\devops) to 56.6\% (\datasicentist).
In fact, \languages\ is the most required hard skill for 9 out of 14 technical roles analyzed.
Even designers must master some sort of languages to better provide their prototypes, such as \textit{css}, \textit{html}, and \textit{javascript}.
However, the concentration of \languages\ is higher for development-based roles, e.g., \mobile\ (39.3\%), \gamedeveloper\ (50.7\%), \fullstack\ (43.5\%), and \frontend\ (49.5\%).
These characteristics make \languages\ the only skills that are significantly mentioned in all roles.

Likewise, development-based roles also demand skills on \libsframeworks, which generally have more than 20\% of participation in this group.
They are specially required for \fullstack\ and \frontend\ (34.1\%, both).

Regarding \osinfrastructure, two roles stand out with more than 50\%: \sysadmin\ (53.0\%) and \devops\ (51.9\%).
Interestingly, \mobile\ appears next with 36.4\%.
This result is explained by the fact that such developers must master mobile operating systems, such as \textit{ios} and \textit{android}.

For \processmethods, \designer\ stands out with 68.7\% of the skills coming from this group.
Indeed, the most mentioned \designer\ skills are related to user interfaces, e.g., \textit{user-experience}, \textit{user-interface}, and \textit{design}. 
Next, \qatest\ and \productmanager\ follow up with 42.9\% and 38.1\%, respectively.
Differently, these roles require hard skills associated with software quality assurance.
For instance, most \qatest\ opportunities require knowledge on automated testing methods.
\productmanager's candidates should have experience in agile development processes such as \textit{scrum}.
By contrast, \processmethods-based skills are not highly demanded for development-based jobs.

Skills on \datasystems\ are more required for \databaseadmin\ and \datasicentist, with 31.1\% and 27.4\%, respectively.
Surprisingly, \devtools-based skills are considerably mentioned only in \qatest\ (22.2\%) and \devops\ (12.6\%).
In fact, tools like \textit{selenium} and \textit{jenkins} are in the top-5 most mentioned ones for this category.

\section{Analyzing Soft Skills}

\begin{figure*}[t!]
    \centering
    \includegraphics[width=0.9\textwidth]{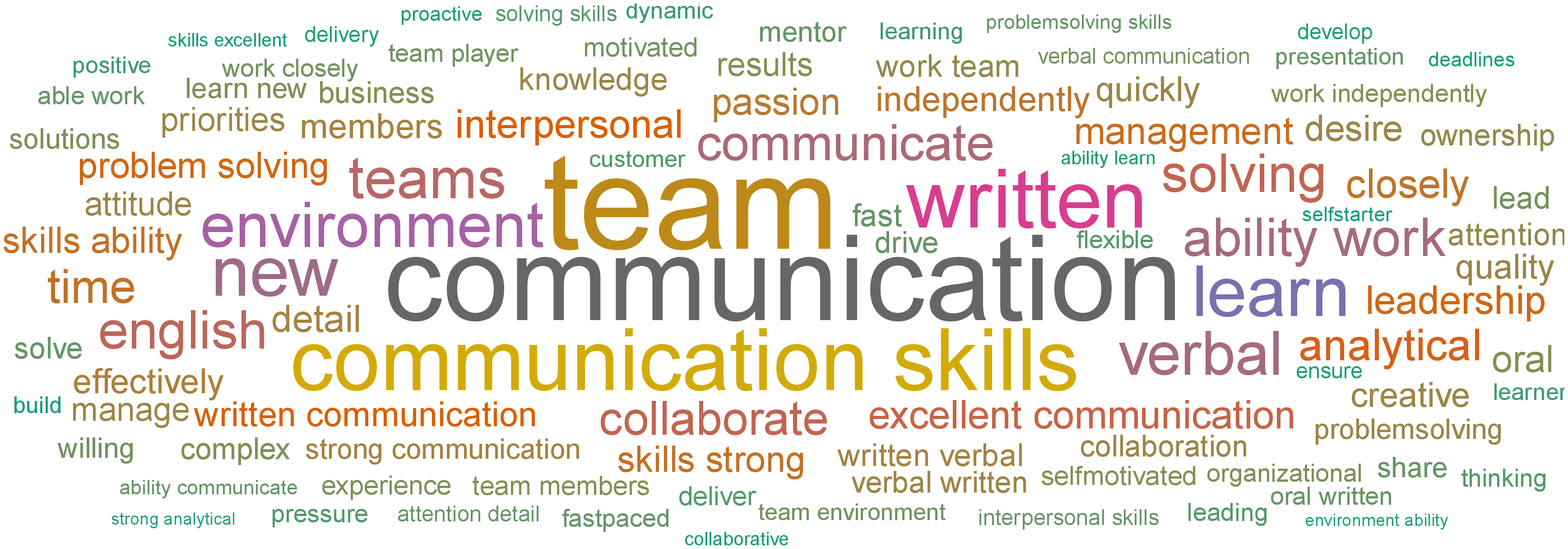}
    \caption{Soft skills word cloud.}
    \label{fig:wordcloud}
\end{figure*}

Unlike hard skills, soft skills are only available on jobs' textual descriptions, i.e., Stack Overflow does not include tags for such skills.
This limitation makes the act of collecting soft skills more complex, as we have to extract them from unstructured text.
We handled this issue by first randomly selecting a sample of 376 opportunities from a subset of 17,756 posts.\footnote{The sample size was determined after specifying a limit of 95\% confidence level and 5\% confidence interval in a well-known sample size calculator, available at \url{https://surveysystem.com/sscalc.htm}.} 
We discarded the remaining 3,212 posts as they were not written in English and, therefore, do not fit into our analysis.
Then, three authors manually annotated sentences that refer to soft skills, such as ``good communication skills'', ``ability to work independently'', etc.
In total, we extracted 1,530 sentences from 315 opportunities.
Next, we generated a word cloud to identify the most common terms among these sentences.
As some soft skills can be described using multiple words (e.g., ``verbal communication''), we adapted the word cloud to consider bigrams as well.

\autoref{fig:wordcloud} depicts the word cloud for the top-100 most frequent terms.
We can observe that \textit{communication} plays a central role among the most required soft skills.
Our sample allows us to conclude that $32\% \pm 5\%$ of the posts in our population (i.e., about one in three posts) mention this skill.
\textit{Collaboration}-based skills also feature a special position in this ranking.
For instance, the word ``team'' appears in at least 22\% of the jobs posts (i.e., $27\% \pm 5\%$).
Lastly, some companies also require experience in \textit{problem-solving} skills, as the words ``analytical'',  ``problem solving'', or ``deliver'' are present in at least one out of ten posts (i.e., $15\% \pm 5\%$).

\section{How do Skills Change Over Time?} 

Litecky et. al.~\citep{Litecky2010} leveraged the most demanded hard and soft skills in the IT industry between July 2007 and April 2008 through an analysis of more than 200,000 online job opportunities.
When we compare their results with ours, we observe some changes regarding hard skills.
For instance, languages such as \textit{C/C++}, \textit{SQL}, and \textit{Java} were frequently mentioned in 2008, as they were listed as major ones in 14 out of 20 positions (70\%).
Likewise, in our study \textit{Languages} is the most requested high-level hard skill in 9 out of 14 developer roles (64\%).
Interestingly, no library or framework appeared among the popular hard skills in 2008.
This represents a contrast with our study, as \textit{Libs \& Frameworks} are the second most demanded hard skill for six roles.
Finally, we observe minor changes in terms of soft skills.
For instance, \textit{leadership} is mentioned in 20\% of the opportunities in 2008.
Similarly, terms like \textit{leadership} (7.6\%), \textit{lead} (4.0\%), and \textit{leading} (3.5\%) are also featured among the most popular soft skills in our study.
However, these numbers should be interpreted with caution, since the studies followed different methodologies.

\section{Conclusion}

For candidates to IT jobs, our study reveals the distribution of the most demanded hard skills for 14 contemporary developer roles. 
For example, people who are applying to a frontend position should describe the \textit{programming languages} and \textit{libraries \& frameworks} they master, since 49.5\% and 34.1\% of the tags used by companies when hiring such developers refer to these hard skills.
Furthermore, the study confirms the importance companies give to soft skills. 
The most requested ones are related to communication, collaboration, and problem-solving.
We plan to extend the study described in this paper in two major directions.
Firstly, by including other jobs portals besides Stack Overflow in our analysis~\citep{Litecky2010}.
Secondly, by investigating machine learning approaches to identify soft skills automatically~\citep{Bastian2014, Sayfullina2018}.
The data used in this study is  available at \url{https://doi.org/10.5281/zenodo.3906955}.



\balance
\bibliographystyle{elsarticle-num}
\bibliography{lib}

\end{document}